\documentclass[pre,twocolumn,superscriptaddress,showpacs,floatfix]{revtex4}
\usepackage{amsmath}    
\usepackage{graphicx}   
\usepackage{verbatim}   
\usepackage{subfigure}    
\usepackage{bbold}
\usepackage{hyperref}   

\begin{document}

\title{The role of chaos in quantum communication through a dynamical dephasing channel}
\author{Gabriela Barreto Lemos}
\affiliation{Instituto de F\'isica, Universidade Federal do Rio de Janeiro, Caixa Postal 68528, Rio de Janeiro, RJ 21941-972, RJ, Brazil
}
\affiliation{Center for Nonlinear and Complex Systems,
Universit\`a degli Studi dell'Insubria, Via Valleggio 11, 22100 Como, Italy}
\author{Giuliano Benenti}
\affiliation{CNISM, CNR-INFM \& Center for Nonlinear and Complex Systems,
Universit\`a degli Studi dell'Insubria, Via Valleggio 11, 22100 Como, Italy}
\affiliation{Istituto Nazionale di Fisica Nucleare, Sezione di Milano,
via Celoria 16, 20133 Milano, Italy}
 
\begin{abstract}
In this article we treat the subject of chaotic environments with few degrees of freedom in quantum communication by investigating a conservative dynamical map as a model of a dephasing quantum channel. When the channel's dynamics is chaotic, we investigate the model's semi-classical limit and show that the entropy exchange grows at a constant rate which depends on a single parameter (the interaction strength), analogous to stochastic models of dephasing channels. We analyze memory effects in the channel and present strong physical arguments to support that the present model is forgetful in the chaotic regime while memory effects in general cannot be ignored when channel dynamics is regular. In order to render the non-chaotic channel forgetful, it becomes necessary to apply a reset to the channel and this reset can efficiently be modeled by application of a chaotic map.  We may then refer to encoding theorems (valid in the case of forgetful channels) to present evidence of a transition from noiseless to noisy channel due to the environment's transition from regular to chaotic dynamics.
\end{abstract}
\pacs{05.45.Mt,03.65.Yz,03.67.-a,05.45.Pq}
\maketitle
\section{Introduction}
The interaction of an open system with its surroundings creates correlations between the system and the environment which may cause suppression of coherence and/or attenuation of entanglement in the system of interest. This phenomenon is called decoherence\cite{zurek} and its implications are that when environmental degrees of freedom are ignored (traced over), quantum information initially present in the state of the system can be lost. Decoherence is considered the key explanation as to why quantum effects are not typically observed in macroscopic systems and is also the biggest obstacle to overcome in the practical implementation of any quantum computation and quantum communication strategy, which all rely on coherence and entanglement resources. Systems with infinitely many degrees of freedom (a collection of harmonic oscillators \cite{caldeira} or spin $1/2$ particles \cite{spin}, for example) have been used to model noisy environments for many years, and predict well dissipation and decoherence in quantum open systems. 

On the other hand,  the environment surrounding any physical system in general possesses non-linearity and randomness which may be relevant to the process of decoherence. With this in mind, many publications (see \cite{kim}-\cite{jaquod}) in the last decade have been  dedicated to understanding the role of chaos in the destruction of coherence and/or entanglement by using quantum systems with few degrees of freedom and chaotic underlying classical dynamics as models for noisy environments. These publications indicate that  the key feature that enables an environment to cause decoherence is the complexity of the environment, may it arise from a large number of degrees of freedom or from chaotic dynamics. Indeed, reference \cite{aguiar2} showed that the effective dynamics of a harmonic oscillator coupled to a chaotic system with two degrees of freedom is analogous to that produced from coupling to a thermal bath in the limit of high temperatures and weak damping \cite{caldeira,caldeira2}. It has also been shown \cite{rossini,bandyop} that the interaction of two non-interacting qubits with single particle chaotic systems (the kicked rotor \cite{rossini} and the kicked top \cite{bandyop}) can produce entropy growth in the qubit subsystem, cause decay of the entanglement initially present between the non-interacting qubits, or create entanglement in initially non-entangled qubits, equivalently to what is expected from the interaction of two qubits with infinitely large canonical environment models.  

The next natural question to address would be: does an environment's transition to chaotic dynamics make it more efficient in causing decoherence? At this point, it is important to say that any environment which has a finite dimensional Hilbert space, can only be effective in causing decoherence for a finite time, and after this time quantum effects in the system can be recovered. In this regard, reference \cite{saraceno} showed that the time over which an environment of finite dimension can be effective is much longer when its underlying classical dynamics is chaotic than when it is in the regular regime (by \textit{regular} we mean non-chaotic). On the other hand, it was shown in \cite{buric} that the rate of decoherence in a qubit pair interacting with a dissipative nonlinear oscillator is enhanced when oscillator dynamics is chaotic rather than regular. Even when compared to a heat bath with infinitely many degrees of freedom, unstable systems (defined by a positive upper Lyapunov exponent) with few degrees of freedom and finite dimension can be more efficient in inducing decoherence, as was shown in \cite{carolina, kohout} for time scales shorter than the saturation time of the finite-dimensional environment. 

In problems involving dynamical evolution and transmission of quantum states through noisy environments, one is led to the mathematical framework of quantum channels. Indeed, communication through quantum noisy channels \cite{NielsenChuang, bookbenenti} is one of the central problems in quantum information theory. It concerns, for example,  the transmission in the presence of noise of an unknown quantum state between two units of a quantum system, like photons being sent through imperfect optical fibers. Other examples of applications of quantum noisy channel models refer to imperfections in a quantum teleportation process or the storage of information in a quantum computer memory (in which case the transmission of the quantum state is through time).

In order to account for errors introduced by noise in the transmission of a quantum state, one assumes that the system of interest $\mathcal{Q}$ initially in the input state $\hat\rho$ interacts with a suitable environment $E$. Before transmitting $\hat\rho$, the information contained in this state is encoded in blocks of $N_q$ qubits which are sent down the channel and then decoded again. The objective is to have a decoded state $\hat\rho'$ at the output of the channel that matches with high fidelity the input state $\hat\rho$. The central question to address for a given noisy channel is: \textit{what is the maximum rate at which quantum information can be transmitted with negligible error in the limit of a large number of channel uses?} The answer is given by the quantum capacity $Q$ of the channel. In the case of memory-less channels (which act in an uncorrelated manner with each qubit), well defined mathematical relations can be used for obtaining the value of $Q$  \cite{lloyd}-\cite{hayden}, but in the more realistic scenario in which memory effects must be taken into account, it is not always straightforward to mathematically formulate the quantum capacity in the limit of infinite channel uses \cite{plenioprl}-\cite{hamedo}. Despite of this, coding theorems for the quantum capacity have been proved for the so-called forgetful channels, for which memory effects decay exponentially in time  \cite{werner}. In particular, the quantum capacity of dephasing channels with memory was studied in references \cite{plenioprl}-\cite{dephasing2}. Dephasing channels, characterized by the existence of preferential orthonormal basis states which are transmitted without errors, will be the focus of the present article.

In the present article we take the subject of chaotic environments with few degrees of freedom a step further in the direction of the quantum computation and quantum communication scenario by proposing a new dynamical model of quantum channel given by single-particle, fully deterministic, conservative dynamical map. 
We present numerical results for the entropy exchange and coherent information of the model and show growth of the entropy exchange at a constant rate in the case of chaotic dynamics while sub-linear growth is observed in the absence of chaos. When the channel is in the chaotic regime,  the constant entropy exchange rate can be varied from $0$ to its maximum achievable value, $1$, by varying the coupling strength, which is in this case the only relevant parameter. We analyze memory effects in the model and observe an interesting difference between regular a chaotic dynamics also in this respect. While numerical results strongly support the conjecture that the present channel model is forgetful~\cite{werner}  in the chaotic regime, this is not the case of regular dynamics, in which memory effects can be significant. In order to render the non-chaotic channel forgetful, one must periodically wipe out memory effects by resetting the quantum channel,  and this reset can be efficiently modeled by the use of a chaotic map. We can then take advantage of encoding theorems and discuss the quantum capacity of the channel in both dynamical regimes. We associate the transition from noiseless ($Q=1$) to noisy channel ($Q<1$) with the transition from regular to chaotic dynamics. 

The article is organized as follows: In section \ref{I} we briefly review the basic quantities and concepts necessary for the characterization of a quantum channel. We then go on to describing the hereby proposed model of a Hamiltonian dephasing channel in section \ref{II}. In section \ref{III} we study the dynamical dephasing channel model in the chaotic regime. We then go on to discuss memory effects and forgetfulness in section \ref{IV}. On the grounds of a forgetfulness conjecture, in section \ref{V} we discuss the capacity of the channel for each dynamical regime. We finish with concluding remarks in section \ref{VI}.

\section{Fundamental concepts of Quantum Information Transmission}\label{I}
In this section we present an overview of the basic concepts and relations concerning 
the transmission of quantum information through a quantum channel contained in a $N_q$ qubit system $\mathcal{Q}$, representing $N_q$ successive uses of quantum channel. The action of the channel is described by a superoperator $\mathcal{E}$ representing a completely positive, trace preserving linear map that transforms the input state $\hat\rho$ of the quantum system $\mathcal{Q}$ into the output state $\hat\rho'$: 
\begin{equation}\label{eq:canal}
 \hat\rho'=\mathcal{E}_{N_q}(\hat\rho)=\rm{Tr}_E\left[\hat U\left(\hat\rho\otimes\hat\omega_0\right)\hat U^\dagger\right],
\end{equation}
where $\hat U$ is the global unitary evolution operator for $N_q$ uses of the channel and the environment's pure initial state is given by $\hat\omega_0=\vert\omega_0\rangle\langle\omega_0\vert$. The corresponding final state of the environment is described by the conjugate map  $\hat\rho^{E'}=\tilde{\mathcal{E}}_{N_q}(\hat\rho)=\rm{Tr}_{\mathcal{Q}}\left[\hat U\left(\hat\rho\otimes\hat\omega_0\right)\hat U^\dagger\right]$. 

Formally, one considers a source of identical quantum systems $\mathcal{Q}$ prepared in an unknown quantum state. The information content of the quantum system $\mathcal{Q}$ is given by the Von Neumann Entropy $S(\hat\rho)=-\rm{Tr}\left[\hat\rho\log_2\hat\rho\right]$~\cite{barnum,schuma}. The reliability of the quantum information transmission is measured by the \textit{entanglement fidelity} \cite{barnum,schuma2}. In order to define this quantity, one must consider a larger quantum system $\mathcal{RQ}$, initially in a pure entangled state $\vert\Psi^{\mathcal{RQ}}\rangle$. The density operator of the quantum system of interest $\mathcal{Q}$ is obtained from $\vert\psi^{\mathcal{RQ}}\rangle$ by taking the partial trace over the reference system $\mathcal{R}$: $\hat\rho=\rm{Tr}_{\mathcal{R}}\left[\vert\psi^{\mathcal{RQ}}\rangle\langle\psi^{\mathcal{RQ}}\vert\right]$. The system $\mathcal{Q}$ is sent through the channel while $\mathcal{R}$ is considered to be isolated from the environment so that the final state of the composite 
system is given by 
\begin{equation}\hat{\rho}^{\mathcal{RQ}'}=(\mathcal{I}^{\mathcal{R}}\otimes\mathcal{E}^{\mathcal{Q}})\left(\vert\psi^{\mathcal{RQ}}\rangle\langle\psi^{\mathcal{RQ}}\vert\right),
\end{equation}
 where $\mathcal{I}$ is the identity superoperator. The entanglement fidelity $F_e$ is then defined as the fidelity between the initial pure state $\vert\psi^{\mathcal{RQ}}\rangle$ and the final state $\hat{\rho}^{\mathcal{RQ}'}$, which will in general be a mixed state:
\begin{eqnarray}
F_e&=&F_e(\hat\rho,\mathcal{E})=F\left(\vert\psi^{\mathcal{RQ}}\rangle,\hat{\rho}^{\mathcal{RQ}'}\right)\\\nonumber
&=&\langle\psi^{\mathcal{RQ}}\vert(\mathcal{I}^{\mathcal{R}}\otimes\mathcal{E}^{\mathcal{Q}})\left(\vert\psi^{\mathcal{RQ}}\rangle\langle\psi^{\mathcal{RQ}}\vert\right)\vert\psi^{\mathcal{RQ}}\rangle.
\end{eqnarray}
The use of an enlarged quantum system $\mathcal{RQ}$ is a mathematical artifice and clearly the entanglement fidelity should not depend on the particular purification $\vert\psi^{\mathcal{RQ}}\rangle$ chosen. Indeed, $F_e$ can be shown to depend only on the initial state $\hat\rho$ of the system of
 interest $\mathcal{Q}$ and on the action of the channel $\mathcal{E}$ \cite{schuma2}.

Another important concept in the study of quantum noisy channels is the \textit{entropy exchange} $S_e$\cite{barnum, schuma2}. This is defined as the entropy that the enlarged system $\mathcal{RQ}$ acquires when $\mathcal{Q}$ interacts with the environment $E$ in the quantum channel. As the environment is initially in a pure state, this is equivalent to the entropy increase of the environment $E$ and is a measure of the entanglement between $\mathcal{RQ}$ and $E$ after subsystem $\mathcal{Q}$ has gone through the channel:
\begin{equation}
S_e=S_e(\hat\rho,\mathcal{E})=S(\hat\rho^{\mathcal{RQ}'})=S(\hat\rho^{E'}),
\end{equation}
where $S(\hat\rho^{E'})$ is the Von Neumann entropy of the final state of the environment. Like the entanglement fidelity, the 
entropy exchange is seen to depend exclusively on the system input $\hat\rho$ and on the channel $\mathcal{E}$, irrespective of the particular purification \cite{schuma2}. 

Before arriving at the relation for the quantum channel capacity, there is one final 
quantity to define: the \textit{coherent information}, $I_c$ \cite{barnum, schuma3}:
\begin{eqnarray} 
 I_c\left(\hat\rho,\mathcal{E}\right) &=& S\left(\mathcal{E}(\hat\rho)\right)-S_e(\hat\rho,\mathcal{E}) \nonumber\\
&=&S(\hat\rho')-S(\hat\rho^{\mathcal{RQ}'}). \label{eq:info}
\end{eqnarray}
When $\mathcal{Q}$ and $\mathcal{R}$ are maximally entangled, $S(\hat\rho)$ is maximal because the input state $\hat\rho$ is maximally mixed. If on top of that the channel is noiseless then the coherent information will be maximal because in this case, $S(\hat\rho')=S(\hat\rho)$ is maximal and $S(\hat\rho^{\mathcal{RQ}'})=0$, since the state $\vert\psi^{\mathcal{RQ}}\rangle$ remains pure after transmission. When $\vert\psi^{\mathcal{RQ}}\rangle$ is not maximally entangled 
or the channel presents noise, smaller values of $I_c$ are obtained. This outlines the fact that the coherent information quantifies the capability of a quantum channel to convey entanglement.

Finally, we have the relation for the \textit{quantum channel capacity} $Q$, taken in the limit of infinite channel uses \cite{lloyd}-\cite{hayden}: 
\begin{eqnarray} \label{capacity}
Q &=& \lim_{N_q\to\infty}\frac{Q_{N_q}}{N_q},\\
Q_{N_q} &=& \max_{\hat\rho} I_c\left(\hat\rho,\mathcal{E}_{N_q}\right),
\end{eqnarray}
where the maximum of $I_c/N_q$ is over all possible input states $\hat\rho$, $N_q\to\infty$.
One must be cautious in using the above relation to calculate the quantum channel capacity $Q$. In general this relation provides only an upper bound for the channel capacity except when it is proved the existence of an encoding that makes it possible to reach this bound. However, in the case of memory-less channels and so-called \textit{forgetful channels} \cite{plenio}, in which memory effects decay exponentially with time, this bound can be reached.

In this article we will study a model of a quantum dephasing channel. The distinguishing property of any pure dephasing channel is that the
total system-environment Hamiltonian $\hat{H}$ commutes with the system Hamiltonian $\hat{H_Q}$,  $[\hat{H}_{\mathcal{Q}},\hat{H}]=0$, implying that the eigenstates ${\vert j\rangle}$ of the $\hat H_{\mathcal{Q}}$  form a preferential orthonormal basis $\{\vert j\rangle\equiv\vert j_1,..,j_{N_q}\rangle$, $j_1,...,j_{N_q}=0,1 \}$. An $N_q$ qubit train in an eigenstate of this basis is not affected by the channel. While this means that dephasing channels are noiseless to transmission of classical information (bits), quantum information encoded in superpositions of basis states may be corrupted. The quantum capacity of dephasing channels with memory was studied in \cite{plenioprl}-\cite{hamedo} and for some specific models in which memory effects die out exponentially fast the quantum capacity was computed.

\section{A dynamical map as a model of a dephasing channel}\label{II}
We investigate a model of dephasing channel described by the total system-environment Hamiltonian
\begin{equation}\label{total}
\hat{H}=\hat{H}_{\mathcal{Q}}+\hat{H}_{K} -\hat F(t)\hat X_E .
\end{equation}
The system Hamiltonian is given by $(\hbar=1)$
\begin{equation}
\hat H_{\mathcal{Q}} =  \sum_{n=1}^{N_q}\hat\sigma_z^{(n)},
\end{equation}
where $\hat\sigma_z^{(n)}$ is Pauli operator for the $n$-th qubit with $\hat{\sigma}_z^{(n)}\vert j_n\rangle=\pm\vert j_n\rangle$. The environment is modeled by the time-dependent Hamiltonian
\begin{equation} \label{kicked}
\hat{H}_K=\frac{\hat p^2}{2}+V(\hat{\theta})\sum_n \delta\left(t-Tn\right), 
\end{equation}
and the system-environment coupling is a product of the environment operator $\hat X_E$ and the system operator
\begin{equation}\label{F}
\hat F(t) = \sum_{n=1}^{N_q}\hat\sigma_z^{(n)} f^{(n)}(t),
\end{equation}
with $f^{(n)}(t)=1$ if qubit $n$ is inside the channel and $f^{(n)}(t)=0$, otherwise.

We will focus our attention on a specific potential $V(\hat{\theta})=-k(\hat{\theta}-\pi)^2/2$
 in eq.(\ref{kicked}). This is the so-called \textsl{quantum sawtooth map} \cite{lichter}-\cite{montangero}, and it is obtained from the quantization of the classical 
sawtooth map 
\begin{eqnarray}\label{mapa}
\left\{\begin{array}{ll} 
p_{i+1}=p_i+k(\theta_i-\pi),\\
\\
\theta_{i+1}=\theta_i+T p_{i+1}\quad(\rm{mod}\: 2\pi),
\end{array}\right.
\end{eqnarray}
describing the classical evolution from the $i$-th kick at time $Ti$ to to the $(i+1)$-th kick at time $T(i+1)$. The variables
$(p,\theta)$ are conjugate action-variables with $0\le\theta<2\pi$. When the map (\ref{mapa}) is rewritten in terms of the rescaled momentum $P=Tp$, the classical dynamics is seen to depend upon the single parameter $K=kT$. For $K>0$ and $K<-4$ the dynamics is completely chaotic, with homogeneous exponential instability. Dynamics is not chaotic for $-4\leq K\leq 0$ (we also use the term \textit{regular} to refer to non-chaotic dynamics)~\footnote{In particular, the sawtooth map is completely integrable for $K=-1,-2,-3,-4$.}. The hypothesis of the Kolmogorov-Arnold-Moser (KAM) theorem \cite{lichter} are not satisfied for such a discontinuous map and the motion is not bounded by KAM tori for any $K\ne 0$. Indeed, the map exhibits normal diffusion in momentum $<\Delta P^2>\approx D(K)t/T$ for any $K>0$, where $D$ is the diffusion coefficient, $t/T$ is the discrete time measured in units of map iterations and $<\cdot\cdot\cdot>$ is the mean performed over an ensemble of particles with momentum $P_0$ and random phases $0\le\theta<2\pi$ \cite{dana}.

The quantum sawtooth map is the quantized version of the classical sawtooth map (\ref{mapa}). The Floquet operator describing the quantum evolution corresponding to one iteration of the sawtooth map is 
\begin{equation}\label{Uk}
\hat{U}_K=\exp\left(-\frac{\imath}{2T }\hat P^2\right)\exp\left(\frac{\imath}{2T}K(\hat{\theta}-\pi)^2\right),
\end{equation}
where we have used the rescaled momentum operator $\hat P=T\hat p=-\imath T \partial /\partial\theta$ ($\hbar=1$). The rescaled momentum $P$ defines an effective Planck constant $[\hat P,\hat \theta]=-\imath T\equiv-\imath\hbar_{eff}$. As the classical sawtooth map is characterized by very different regimes, so is its quantum version, but while classical dynamics depends on the single parameter $K=kT$, quantum dynamics depends independently on each parameter $k$ and $T$. The quantum sawtooth map presents important physical phenomena, like dynamical localization, for example \cite{montangero}.

The map (\ref{mapa}) can be studied on the cylinder $[P\in (-\infty,\infty)]$, or can be closed to form a torus of length $2\pi L$, where $L$ is an integer. In this paper we consider sawtooth map with the phase space closed on the torus $0\leq\theta < 2\pi$, $-\pi\leq P< \pi$.  In this case, the number of quantum levels $N$ is related to the effective Planck constant by $\hbar_{eff}=T=2\pi/N$, and the sawtooth map describes a conservative dynamical model of an environment with two degrees of freedom. We would just like to make a note that in order to remove time reversal invariance $P\rightarrow -P,\theta\rightarrow 2\pi-\theta$ of the model, one must apply the transformation $P\rightarrow P +T\phi_0,\theta\rightarrow\theta+T\theta_0$, where $\phi_0$ plays the role of an Aharonov Bohm-flux \cite{felix}. In the following we use the values $\phi_0=\theta_0=\sqrt{2}/5$. Without doing this transformation, the quantum sawtooth map cannot attain maximal Von Neumann Entropy $S_{max}=\log_2 N$, associated with a maximally mixed state in a $N$ dimensional Hilbert space.  In the following sections we refer to the model's semi-classical limit $\hbar_{eff}\to 0$, obtained by increasing the Hilbert Space dimension $N\to\infty$, while doing $k\to \infty$ so as to maintain the classical parameter $K=kT$ fixed. 

In computing the numerical results presented in this paper, we used a kicked interaction Hamiltonian
\begin{eqnarray}\label{eq:interaction}
 \hat{H}_I&=&-\hat F(t)\hat X_E,\\
\hat X_E&=&\hat X_E(\hat\theta)=\frac{\eta (\hat{\theta}-\pi)^2}{2}\sum_n\delta\left(t-Tn\right),\label{int1}
\end{eqnarray}
so that the interaction could be seen as a re-scaling of the kicking parameter $K$ (the system operator $\hat F$ is given by eq.\ref{F}). We would like to point out that other 
kicked or continuous environmental interaction operators $\hat X_E(\hat\theta,\hat{p})$~\footnote{Specifically, we ran tests with $\hat X_E=\eta\cos{\hat\theta}\sum_n\delta\left(t-Tn\right)$, $\hat X_E=\eta \sin{\hat{p}}$,  $\hat X_E=\eta\hat{p}$, for example.} in eq.(\ref{eq:interaction}) were analyzed and qualitatively 
equivalent results were obtained in each case, indicating that the specific form of the operator $\hat X_E(\hat\theta,\hat{p})$ and 
the fact that it is or not continuous in time does not alter the main features of this quantum dephasing channel model.

We call $\tau_p$ and $\tau$ respectively the time each qubit remains in the channel and the time interval 
between the entrance of two consecutive qubits. Initially $(t=0)$ the system and environment are not entangled, and we are interested in the total evolution 
time $\tau_{N}=(N_q-1)\tau+\tau_p$ corresponding to transit time for an $N_q$ qubit train. We will conveniently set $\tau_p=T$, 
where $T$ is the time interval between consecutive kicks in $\hat{H}_K$. This means that each qubit is in the 
channel for exactly one iteration of the quantum map. We will also set $\tau=n_0 T$, where $n_0$ is an integer, so that the 
total time between consecutive entries in the channel is a multiple of the time each qubit is in the channel. 

\section{The dynamical dephasing channel model in the chaotic regime}\label{III}
 In this section we will study the entropy exchange and the coherent information in the present dephasing channel model in the chaotic regime. It is important to point out that in the current section we consider that memory effects are ignorable and indeed this is a reasonable assumption in the chaotic regime, as will be discussed in detail in the next section \ref{IV}.  

We begin by considering a pure initial global state in which the qubits are disentangled from one another and from the initial pure state of the environment $\hat\omega_0=\vert\omega_0\rangle\langle\omega_0\vert$:  
\begin{eqnarray}\label{initial}
\vert\Psi(t=0)\rangle\langle\Psi(t=0)\vert&=&\vert\omega_0\rangle\langle\omega_0\vert\otimes\hat\rho;\\
\vert\omega_0\rangle &=& \sum_P c_P\vert P\rangle,
\end{eqnarray}
where $c_P$ are random coefficients generated according to the Haar measure \cite{haar}, with $\sum_P \vert c_P\vert^ 2=1$ and $\vert P \rangle$ are eigenstates of the environment momentum operator. We are interested in the evolution due to the passage of an $N_q$ qubit train, which corresponds to the time $\tau_N=(N_q-1)\tau+T
=((N_q-1)n_0+1)T$, after which the final state $\hat\rho'$ of the $N_q$ qubit system is the given by
\begin {equation}
\hat\rho'=\rm{Tr}_E\left\{\hat{U}(\tau_N)(\hat\rho\otimes\hat\omega_0)\hat{U}^\dagger(\tau_N)\right\},
\end{equation}
where 
\begin{equation} \label{evol}
\hat{U}(\tau_N)=\hat{\mathcal{T}}\exp\left(-\imath\int_0^{\tau_N} dt \hat{\tilde{H}}(t)\right)
\end{equation}
and $\hat{\tilde{H}}(t)$ is the Hamiltonian (\ref{total}) expressed in the interaction picture with respect to the system Hamiltonian $\hat {H}_{\mathcal{Q}}$ and $\hat{\mathcal{T}}$ is a time ordering operator. 

Because $\hat H_{\mathcal{Q}}$ commutes with the total Hamiltonian $\hat H$, the dynamics preserves the basis ${\vert j\rangle =\vert j_1, ...,j_{N_q}\rangle}$ formed by the eigenstates of $\hat H_{\mathcal{Q}}=\sum_{n=1}^{N_q}\hat\sigma_z^{(n)}$,
\begin{equation}
\hat U(t)\left(\vert j \rangle\vert\phi\rangle\right)=\left(\hat{\openone}\otimes \hat U_j(t)\right)\vert j \rangle\vert\phi\rangle,
\end{equation}
where ${\vert \phi\rangle}$ is a complete orthonormal basis of the environment and $\hat U_j(t)=\langle j\vert\hat U(t)\vert j\rangle$ are conditional evolution operators acting only on environmental degrees of freedom ($\hat{\openone}$ is the identity operator).  So the elements of the final reduced density matrix of the system are given by
\begin{eqnarray}
(\hat\rho')_{jl}&=&\langle j\vert \rm{Tr}_E \left\{ \hat U(\tau_N) (\hat\rho\otimes\hat\omega_0 )\hat U ^\dagger (\tau_N) \right\} \vert\textit{l}\rangle\nonumber\\
&=&(\hat\rho)_{jl}\langle\omega_l\vert\omega_j\rangle,\label{off}
\end{eqnarray}
where $\vert\omega_j\rangle\equiv\hat U_j\vert\omega_0\rangle$ are conditional states of the environment. The overlap between different conditional environmental states  $\langle\omega_l\vert\omega_j\rangle ,(l\neq j)$ determines how the system's off-diagonal matrix elements change due to interaction with the environment, while unitarity $\langle\omega_j\vert\omega_j\rangle=1$ guaranties the preservation of populations $(\hat\rho_{jj})$. It is therefore clear that diagonal input states do not change during evolution, 
$S[\mathcal{E}_{N_q}(\hat\rho)]=
S(\hat\rho) $, implying that in this case the coherent information (\ref{eq:info}) can be easily computed from the entropy exchange 
$S_e=S[\tilde{\mathcal{E}}_{N_q}(\hat\rho)]$ and the input state $\hat\rho$. Moreover, if memory effects can be ignored, the coherent information in any pure dephasing channel is maximum for the completely unpolarized input state $\hat\rho_{un}\equiv \left(1/2^ {N_q}\right)\hat\openone_1^{\otimes N_q}$, where $\hat\openone_1$ is the identity operator acting on the single-qubit Hilbert space, and therefore this diagonal state is conveniently the input state one should use in the optimization (\ref{capacity}) to find the channel capacity.  

With this input state, we compute the entropy exchange $S_e$ of the quantum channel with fixed 
coupling constant $\eta$ and time scale $\tau=T$ (\textit{i.e.} there is no time delay between the exit of a qubit and the entrance of the next qubit into the channel). Because the input state $\hat\rho_{un}$ is left unchanged by the channel, the coherent information $I_c$ can be trivially computed from $S_e$  by the relation $I_c=N_q-S_e$, obtained from substituting $S(\hat\rho')=S(\hat\rho_{un})=N_q$ in eq.(\ref{eq:info}). Therefore, every time we mention the entropy exchange below, we are also implicitly making statements about the coherent information. 

\begin{figure}[t]
\centering
\includegraphics[angle=270,scale=0.34]{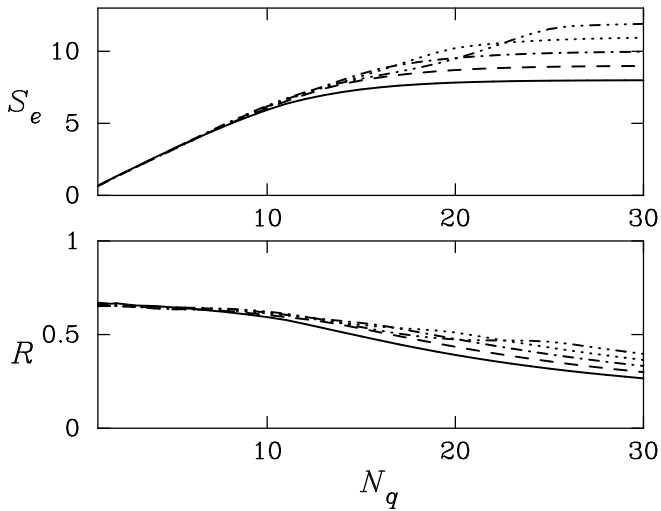}
\caption{The entropy exchange $S_e$ (upper graph) and the entropy exchange rate $R=S_e/N_q$ (lower graph) for the quantum sawtooth map 
as a function of the
number of channel uses $N_q$ at $K=\sqrt{2}$, $\eta=0.3$ and $\tau=T$. The various curves correspond to different values of 
environment's dimension $N=2\pi/T=2^{8}$, $2^{9}$, $2^{10}$, $2^{11}$, $2^{12}$ from bottom to top. The larger the environment's dimension, the more channel uses it takes to saturate the entropy exchange, 
which otherwise tends to grow at a constant rate. \label{fig:limit}} 
\end{figure}

We investigate the semi-classical limit of the model by increasing the dimension $N=2\pi/T$ of the sawtooth map while keeping constant the classical parameter 
$K$ at a value which corresponds to 
classical chaotic regime. This numerical analysis can be seen in Fig.~\ref{fig:limit} where we present the entropy exchange $S_e$ as a function of the number of qubit uses $N_q$ (upper graph) and the entropy exchange rate 
$R=S_e/N_q$  as a function of $N_q$ (lower graph) for different values of the effective Planck constant $\hbar_{eff}=T$. The entropy exchange increases at a constant rate until the channel saturates due to its finite dimensional Hilbert space. As $\hbar_{eff}$ decreases, the channel saturates after an 
ever larger number of channel uses and in the semi-classical limit we expect a constant entropy exchange rate 
$R$  even when the number of channel uses becomes very large. Indeed, if one wishes to simulate the passage of a fixed number $N_q$ of qubits then one must choose the Hilbert space dimension $N=2\pi/\hbar_{eff}$ accordingly so as to not reach the saturation of the basis within the transit time of the qubit train. The larger the number of qubits in the train, the more quantum levels one must consider so that the quantum map is a good dynamical model of a dephasing channel.

In a stochastic model of a memory-less dephasing channel~\cite{puredephasing} a single parameter $g$ controls the entropy exchange rate $R$ and we will see that the present dynamical model of dephasing channel in the chaotic regime also has a single dephasing parameter, namely the coupling strength $\eta$. We begin with an analytical approach to a single channel use $\mathcal{E}_1$, which is actually sufficient to describe any number of uses of a dephasing channel because we are assuming memory effects to be ignorable $\mathcal{E}_{N_q}=\mathcal{E}_1^{\otimes N_q}$. Let us first consider a single use of a pure dephasing channel. Each qubit transmitted through the memory-less dephasing channel with input state $\hat\rho$ has an output state given by:
\begin{eqnarray}\nonumber
\hat\rho=\left(\begin{array}{cc}
\rho_{00}&\rho_{01}\\
\rho_{10}&\rho_{11}
\end{array}\right)\rightarrow\hat\rho'=\left(\begin{array}{cc}
\rho_{00}&(1-g)\rho_{01}\\
(1-g)\rho_{10}&\rho_{11}
\end{array}\right),
\end{eqnarray}
where $g$ is a dephasing parameter that can vary from $g=0$ (no dephasing occurs) to $g=1$ (the qubit loses all its coherence in the transmission). We consider the maximally entangled state $\vert\psi^{\mathcal{R}\mathcal{Q}}\rangle=(\vert 00\rangle+\vert 11\rangle)/\sqrt{2}$ of the larger quantum system $\mathcal{R}\mathcal{Q}$ from which the maximally mixed qubit input state $\hat\rho=\openone/2$ is obtained by tracing over the reference system $\mathcal{R}$. Transmission of system $\mathcal{Q}$ through the memory-less dephasing channel thus yields
\begin{eqnarray}\label{single}
\hat\rho^{\mathcal{RQ}}&=&\frac{1}{2}\left(\begin{array}{cccc}
1&0&0&1\\
0&0&0&0\\
0&0&0&0\\
1&0&0&1\\
\end{array}\right)\nonumber\\
\rightarrow\hat\rho^{\mathcal{RQ}'}&=&\frac{1}{2}\left(\begin{array}{cccc}
1&0&0&(1-g)\\
0&0&0&0\\
0&0&0&0\\
(1-g)&0&0&1\\
\end{array}\right).
\end{eqnarray}
The entropy exchange rate for a single channel use $N_q=1$ is just the Von Neumann entropy of the global quantum system  $R=S_e=S[\hat\rho^{\mathcal{RQ}'}]$ and is obtained from (\ref{single}):
\begin{equation}\label{chanrate}
R=-\left(\frac{2-g}{2}\right)\log_2\left(\frac{2-g}{2}\right)-\left(\frac{g}{2}\right)\log_2\left(\frac{g}{2}\right).
\end{equation}
In fact, this is directly the quantum capacity $Q$ of the dephasing channel because this channel is \textit{degradable}~\cite{puredephasing}, a term that refers to channels in which the final state of the environment can be reconstructed from the final state $\hat\rho'$ of the system. In this case, the regularization in (\ref{capacity}) is unnecessary, and therefore the quantum capacity is given by the 'one-shot' formula $Q=Q_1$. It is thus clear that in a stochastic model of a dephasing channel, the entropy exchange rate $R$, and consequently the channel capacity, depend on a single dephasing parameter $0\le g\le 1$. We will now see that as long as we restrict ourselves to the chaotic regime in the present dynamical model of dephasing channel, the resulting entropy exchange rate also depends on a single parameter $\eta$. 

\begin{figure}[b]
\centering
\includegraphics[angle=270,scale=0.34]{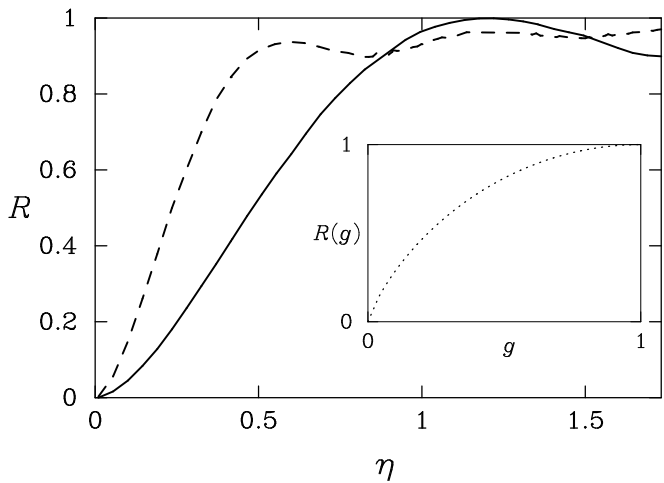}
\caption{The constant entropy exchange rate $R=S_e/N_q$ in the semi-classical limit, as a function of the coupling constant $\eta$ for fully chaotic underlying classical dynamics $K=\sqrt{2}$ of the sawtooth map. The dashed line shows the curve obtained when the interaction is discontinuous in time and given by $\hat X_E(\hat\theta)=\frac{\eta (\hat{\theta}-\pi)^2}{2}\sum_n\delta\left(t-Tn\right)$, while the full line shows the curve obtained when the interaction is continuous with the coupling operator given by $\hat X_E=\eta \sin{\hat{p}}$. For relatively small values of the coupling parameter, the functional dependency of $R$ on $\eta$ is universal for all forms of interaction, but in the non-perturbative regime one can observe a non-monotonous behavior which depends on the specific interaction operator. One also notices that the continuous interaction (full line) takes the entropy exchange rate to its maximum value $R=1$ for smaller coupling strengths when compared to the kicked interaction (dashed line). Identical curves are obtained for other values of $K$, provided one remains in the chaotic regime. The inset shows the entropy exchange rate $R(g)$ of a memory-less dephasing channel as a function of the single dephasing parameter $g$. The present dephasing channel model is seen to be analogous to a pure dephasing quantum channel with the coupling constant $\eta$ playing the role of the dephasing parameter $g$ in the stochastic model.\label{fig:acop}} 
\end{figure}

In order to understand the role of the coupling strength $\eta$ to the entropy exchange growth rate, we will explicitly write out the conditional evolution operators $\hat U_j(t)=\langle j\vert\hat U(t)\vert j\rangle$ acting on environmental degrees of freedom. By making use of the natural time discretization in the problem, we can write
\begin{equation}
\hat{U}_j(\tau_N) =\hat{U}_j^{(a)}(N_q)\hat{U}^{(b)}\hat{U}_j^{(a)}(N_q-1)...\hat{U}^{(b)}\hat{U}_j^{(a)}(1), 
\end{equation}
where, $\hat{U}_j^{(a)}(n)$ refers to the passage of the $n$-th qubit, while $\hat{U}^{(b)}$ refers to the evolution from just after 
the exit of one qubit to just before the entrance of the next qubit in the channel:
\begin{eqnarray}\nonumber
\hat{U}_{j}^{(a)}(n)&=&\exp\left(-\frac{\imath}{2 T}\hat P^2\right)\exp\left(\frac{\imath}{2 T}\left(K-\eta T j_n\right)(\hat{\theta}-\pi)^2\right),\\
\hat U^{(b)}&=&\left(\hat{U}_{K}\right)^{n_0-1},\nonumber
\end{eqnarray}
and $\hat{U}_{K}$ is written in (\ref{Uk}). When the string of qubits passes through the channel, the overlap $\langle\omega_l\vert\omega_j\rangle$, $l\neq j$, determines the decay of the off-diagonal elements of the system's reduced density matrix, as can be seen in eq.(\ref{off}). This overlap  between different states of the environment is essentially the so-called \textit{fidelity} of the chaotic environment, $F=\vert\langle\omega_l\vert\omega_j\rangle\vert ^2$, a widely studied subject in the field of quantum chaos (see, for example, \cite{prosen} and references therein). For illustrative purposes, let us consider a single use of the channel
\begin{eqnarray}
(\hat\rho')_{jl}&=&(\hat\rho)_{jl}\langle\omega_0\vert\hat U_l(\tau)^\dagger\hat U_j(\tau)\vert\omega_0\rangle \label{fidelity}\\
&=&(\hat\rho)_{jl}\langle\bar\omega_0\vert e^{\frac{\imath}{2}\eta (l_1-j_1)(\hat{\theta}-\pi)^2}\vert\bar\omega_0\rangle,
\end{eqnarray}
where $\vert\bar\omega_0\rangle=\exp\left(-\imath\hat P^2/2T\right)\vert\omega_0\rangle$. We can see here the essential role played by the coupling strength $\eta$ in the evolution of the system's reduced density matrix. Indeed, the dashed curve in fig.(\ref{fig:acop}) shows that by varying $\eta$ with the parameter $K$ fixed at a value corresponding to chaotic dynamics of the environment, the entropy exchange rate $R=S_e/N_q$ can increase from $0$ to its maximum possible value $1$. This is analogous to the role played by the dephasing parameter $g$ in the entropy exchange rate $R=R(g)$ of the stochastic dephasing channel model (eq.(\ref{chanrate})), shown in the inset of the same figure. The comparison of the two curves also evidences the non-monotonous behavior of $R$ as a function of $\eta$, as opposed to the monotonously growing curve given by eq.(\ref{chanrate}). In the non-perturbative coupling regime of the dephasing channel model provided by the kicked chaotic map, the entropy exchange rate as a function of $\eta$ presents oscillations which depend on the specific form of the coupling operator $\hat X_E$ in (\ref{eq:interaction}), while for relatively small values of the coupling parameter, the functional dependency of $R$ on $\eta$ is universal for all forms of interaction. To illustrate this, in addition to plotting $R$ as a function of $\eta$ for the coupling operator defined in eq.(\ref{int1}) (dashed line in the main graph of fig.(\ref{fig:acop})), we also show the equivalent curve in the case of a continuous coupling operator $\hat X_E=\eta\sin{\hat{p}}$ (full line). One also notes that in comparison to the kicked coupling, this continuous interaction takes the entropy exchange rate to its maximum value $R=1$ for smaller values of the coupling strength $\eta$. This dependency on the specific coupling can also be seen if one computes the off-diagonal matrix elements in eq.(\ref{off}) in the non-perturbative regime, while for small values of $\eta$, the overlap $\langle\omega_l\vert\omega_j\rangle$, ($l\neq j$) is seen to decay in time  independently of the form of the coupling operator $\hat X_E$  and at a rate $\Gamma\propto\eta^2$, corresponding to the Fermi Golden Rule in fidelity studies~\cite{prosen}.  

Two final remarks can be made about the results hereby presented: First of all, we point out that the classical kicking parameter $K$ does not affect the entropy exchange rate as long as its value corresponds to underlying chaotic dynamics, as will be discussed in section \ref{V}. Secondly, in all the above graphs, we used $\tau=\tau_p$ , but in fact, we also investigated $\tau=n_0\tau_p$ for different values of $n_0$ which corresponds to letting the environment evolve freely between the passage of consecutive qubits. The results we obtain in these cases are identical to those obtained for $n_0=1$. This is a strong indication that indeed memory effects can be ignored in this quantum channel model in the chaotic regime and we dedicate the next section \ref{IV} to sustain this argument as well as to analyze memory effects when the sawtooth map presents regular dynamics.

\section{Memory effects and forgetfulness}\label{IV}
In the calculation of the quantum channel capacity $Q$, the relation (\ref{capacity}) is not always applicable in the case of quantum channels with memory, also called \textit{correlated quantum channels}. In these 
channels,  the transformation $\mathcal{E}_{N_q}$ corresponding to $N_q$ uses of the channel cannot be written as an $N_q$-fold tensor product of the single use channel $\mathcal{E}_1$:
\begin{equation}
\mathcal{E}_{N_q}\neq\mathcal{E}_{1}\otimes\mathcal{E}_{1}\otimes ..\otimes\mathcal{E}_{1}.
\end{equation}
If, however, memory
effects decay exponentially with time, characterizing the so-called \textit{forgetful channels}~\cite{werner}, the relation (\ref{capacity}) may correctly represent the quantum capacity of the memory channel.
 
An operational approach to forgetful channels usually consists of a double-blocking strategy \cite{werner} which should point out when
a memory channel can be mapped into a memory-less one with negligible error. One considers blocks of $M=N_q+L$ uses of the channel 
and does the coding and decoding for the first $N_q$ uses, ignoring the remaining $L$ uses. This means that the passage of $N_q$ qubits through the channel is followed by a time $\tau_L$ corresponding to $L$ idle uses.  If $M$ uses 
of such blocks are considered, the completely positive trace preserving map $\mathcal{E}_{M(N_q+L)}$ can be approximated by 
the memory-less setting $(\mathcal{E}_{(N_q+L)})^{\otimes M}$ with arbitrarily small error when the correlations among 
different blocks decay fast enough during the $L$ idle uses. This property can be expressed by the inequality
\begin{equation}\label{forgetful}
 \|\mathcal{E}_{M(N_q+L)}(\hat\rho) -\mathcal{E}_{(N_q+L)}^{\otimes M}(\hat\rho)\|\leq h(M-1)c^{-L},
\end{equation}
for any input state $\hat\rho$, where $\| \hat{A} \|=\rm{Tr}\sqrt{\hat{A}^\dagger \hat{A}}/2$ is the trace distance, $c>1$ and $h$ depend on the memory model. This relation is not a necessary but a sufficient condition for forgetfulness and it essentially means that although the error committed by replacing the memory channel with 
its memory-less counterpart grows with the number of blocks $M$, it goes to zero exponentially fast with the number $L$ of idle uses in a single block. 
\begin{figure}[t]
\centering
\includegraphics[scale=0.33,angle=270]{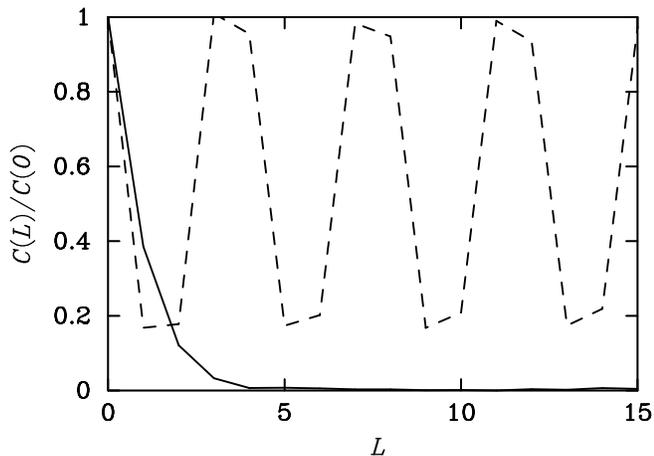}
\caption{The normalized classical auto-correlation function $C(L)/C(0)$, $C(L)=\vert\langle G(L) G(0)\rangle-
\langle G(L)\rangle\langle G(0)\rangle\vert$ as a function of the number of map iterations for two different values of the
classical parameter $K=-\sqrt{2}$ (dashed line) and $K=\sqrt{2}$ (full line). When $K>0$ the classical dynamics is completely chaotic and the auto-correlation function decays
exponentially fast. \label{classcorr}} 
\end{figure}

Returning to the present model of a dephasing channel, when the environment's dynamics is chaotic ($K>0$ or $K<-4$),
our numerical investigations show that changing the ratio $n_0=\tau/T$ between the time separating consecutive qubit entries in the channel
and each qubit's transit time does not alter the entropy exchange as a function of the number of channel uses $N_q$. This indicates
that memory effects may not be important in this dynamical regime of the environment. However, we would like to investigate the
transition from regularity to chaos in our model, and this obliges us to take into account possible memory effects in both chaotic and regular regimes.    

We begin our analysis by investigating the classical auto-correlation function $C(L)=\vert\langle G(L) G(0)\rangle-
\langle G(L)\rangle\langle G(0)\rangle\vert$ , where the function $G$ is associated to the 
operator $\hat G$ acting on environmental degrees of freedom and present in the qubit-environment interaction Hamiltonian (\textit{e.g.} $\hat G=\left(\hat\theta-\pi\right)^2$, for the interaction 
(\ref{eq:interaction})).  In figure \ref{classcorr} we present numerical calculations of the classical auto-correlation function for $L$ iterations of the sawtooth map (\ref{mapa}) where it can be seen that in the case of fully chaotic dynamics, the classical auto-correlation function  decays exponentially fast. This, however, is not true in the case of regular dynamics, in which the auto-correlation function oscillates around a non-zero value, always returning to it's maximum value.  This means that while we can expect that the channel has ignorable memory effects and is forgetful in the chaotic regime, we would not expect forgetfulness when the channel is in the regular dynamical regime.

Based on these results, we must rethink the use of the double blocking strategy described above when the quantum channel is not chaotic. In this case, we must wipe out memory effects by applying a reset of the channel after the passage of a string of $N_q$ qubits and before the next train of qubits. An efficient way to do this reset is by means of a chaotic map. With this in mind, even when the channel dynamics is regular during the $N_q$ qubit train transmission, we use the double blocking strategy always considering fully chaotic dynamics during the $L$ idle uses of the channel. The use of chaotic dynamics during the $L$ idle uses then kills any correlations that might have built up during the $N_q$ channel uses due to regular dynamics in these time periods. By doing this, we aim to render the channel forgetful also when $-4\leq K\leq 0$ and then refer to encoding theorems to conjecture about the channel capacity in all dynamical regimes. 

At this point, it is important to say that exponential decay of the channel auto-correlation function can indicate forgetfulness but by itself does not guarantee it. To further support our forgetfulness conjecture in the case of chaotic dynamics and to show the efficiency of the reset method in rendering forgetful also the non-chaotic channel, we numerically compute inequality (\ref{forgetful}) for the special case of blocks of length $N_q=1$. The triangular inequality guarantees that it is sufficient to prove inequality (\ref{forgetful}) in the case of two blocks $M=2$, so we  consider $M=2$, but limit ourselves (for numerical purposes \footnote{In order to investigate inequality (\ref{forgetful}), one must calculate the maximum trace distance for many different initial states $\hat\rho$ of the qubit system $\mathcal{Q}$. For each initial state $\hat\rho$ we calculate the final state $\mathcal{E}_{2(N_q+L)}(\hat\rho)$ after two successive blocks of $N_q+L$ uses and also the final state $\mathcal{E}_{(N_q+L)}^{\otimes 2}(\hat\rho)$ in which the environment state is reset after the first block of $N_q+L$ uses. In the case of $N_q=1$ we would have $15$ parameters defining the initial state. Conveniently, by convexity, it is sufficient to maximize over pure states (see B. Rosgen, arXiv:$0909.3930$ ($2009$)). Moreover, phases do not affect the final trace distance, so this brings us down to only $3$ free parameters of the qubit initial state to explore in our simulation. The large number of free parameters for $N_q>1$ is the reason why we limit our simulations to single channel uses.}) to single channel uses $N_q=1$ separated by idle times $L\tau_p$ in which chaotic dynamics is always considered. We calculate numerically both the final output state in which memory effects are taken into account and that in which memory effects are ignored completely. By randomly choosing a large number of input states $\hat\rho$, we plot the maximum trace distance as a function of the number $L$ of idle uses and prove the inequality (\ref{forgetful}) both for the case in which the environment's dynamics is chaotic and the case in which it is regular during the passage of the qubits, even though the dynamics during the $L$ idle uses is always fully chaotic ($K=\sqrt{2}$).

\begin{figure}[t]
\centering
\includegraphics[scale=0.33,angle=270]{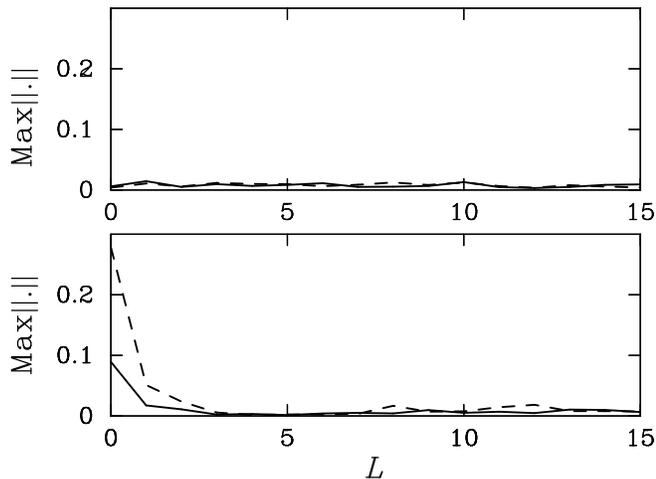}
\caption{The maximum trace distance $\|\mathcal{E}_{2(1+L)}(\hat\rho) - \mathcal{E}_{(1+L)}^{\otimes 2}(\hat\rho)\|$ between the final output state $\mathcal{E}_{2(1+L)}(\hat\rho)$ in which memory effects are taken into account and that in which memory effects are ignored completely $\mathcal{E}_{(1+L)}^{\otimes 2}(\hat\rho)$ as a function of the number $L$ of idle uses. The full line corresponds to chaotic dynamics of the channel $K=1.43$ and the dashed line corresponds to regular dynamics $K=-1.64$. In the upper graph a random initial state of the environment was considered and in the lower graph the initial state is an eigenstate of the environment momentum operator. One can see that inequality (\ref{forgetful}) is fulfilled even when phase-space correlations are present in the initial state of the environment and dynamics is regular (lower graph, dashed line). In these numerical calculations we used coupling constant $\eta=0.3$ and environment Hilbert space dimension $N=2^{12}$. The trace distance displays residual fluctuations of the order of $\sqrt{\hbar_{eff}}$.\label{forg}} 
\end{figure}

In figure (\ref{forg}) we present two different graphs corresponding to different initial states of the sawtooth map. The upper graph was generated considering, as before,  a random initial state of the environment $\vert\omega_0\rangle=\sum_P c_P\vert P\rangle$, where $c_P$ are random coefficients generated according to the Haar measure~\cite{haar}). One can see from this graph that the maximum trace difference when the environment's dynamics is regular is of the same order of that in the case of chaotic dynamics even when there are no (chaotic) idle uses between the passage of the two qubits $L=0$. In fact, the application of a chaotic map is an efficient way to generate a random state \cite{random1,rossini2} and so a random initial state is equivalent to having many chaotic idle uses before the passage of the first qubit and there are essentially no phase-space correlations present in this initial state. As we are considering only single qubit transmissions $N_q=1$, the channel does not have time to build up correlations even for regular dynamics of the sawtooth map during the passage of the qubits. Because of this, regular and chaotic dynamics yield the same results, namely that even for $L=0$ the output state $\mathcal{E}_{2(1+L)}$ is essentially indistinguishable from the memory-less setting $\mathcal{E}_{(1+L)}^{\otimes 2}$. On the other hand, it is expected that should we perform the same calculation for a larger number of channel uses in each block $N_q\gg 1$, regular dynamics during these uses would result in the build up of correlations and a non-zero number of chaotic idle uses $L>0$ would be necessary to break down these correlations between blocks of uses and wipe out memory effects in the channel. In order to show how the chaotic idle uses can wipe out phase-space correlations, we use an initial state of the sawtooth map that is an eigenstate of the environment momentum operator $\vert \omega_0\rangle=\vert P_0\rangle$ and plot in the lower part of fig.(\ref{forg}) the maximum trace distance between $\mathcal{E}_{2(1+L)}$  and $\mathcal{E}_{(1+L)}^{\otimes 2}$ as a function of the number $L$ of idle uses. This graph shows that even if phase-space correlations are present in the environment, the sufficient condition for forgetfulness (\ref{forgetful}) is satisfied both in the case of chaotic and in the case of regular dynamics of the sawtooth map because the difference between taking into account memory effects of the channel and considering a memory-less setting goes to zero sufficiently fast with the number $L$ of chaotic idle uses in a single block. The trace distance then presents only residual fluctuations that go to zero with the effective Planck constant. 

Based on the exponential decay of classical auto-correlation functions $C(L)$ and from the observation that changing the ratio $n_0=\tau/T$ does not alter the coherent information per channel use in the chaotic scenario, and supported on the results for the double blocking strategy in the case of two single uses, we conjecture that the present channel model can be considered forgetful, meaning that inequality (\ref{forgetful}) holds for any $N_q$ and that relation (\ref{capacity}) can therefore be used to calculate the channels capacity $Q$ both in the case of regular and chaotic regimes of the sawtooth map. This is the subject of the next section, where we seek a connection between the channel capacity and the classical chaoticity parameter $K$.

\section{The transition from noiseless to noisy quantum channel}\label{V}
\begin{figure}[b]
\centering
\includegraphics[scale=0.33,angle=270]{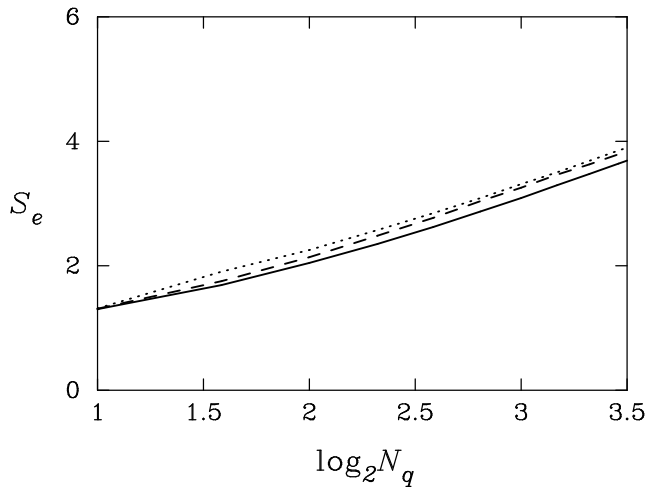}
\caption{The entropy exchange $S_e$ as a function of $\log_2N_q$ for different values 
of the classical parameter $K=-1.8$ (full line), $K=-2.3$ (dashed line), $K=-2.8$ (dotted line), all corresponding to regular dynamics. One can see that the entropy exchange grows logarithmically with the number of channel uses $S_e\propto\log_2 N_q$. This implies that in the regular regime the channel is asymptotically noiseless. Other parameter values are $N=2\pi/T=2^{12}$ and $\eta=0.3$.}\label{kpic}
\end{figure}

\begin{figure}[t]
\centering
\includegraphics[scale=0.34]{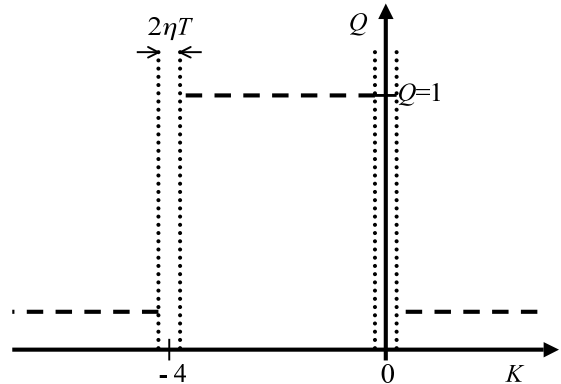}
\caption{Qualitative figure to illustrate the transition from a noiseless $Q=1$ to a noisy $Q<1$ quantum
channel while varying the classical parameter value $K$ through the transition from regular dynamics $-4\le K\le 0$, to chaos $K<-4$ or $K>0$.}\label{fig:trans}
\end{figure}

Assuming forgetfulness of the present channel dephasing model, in this section we will analyze the dependence of the quantum capacity of the quantum channel in the semi-classical limit 
with the underlying classical dynamics of the environment. For $K>0$ or $K<-4$, associated with classical chaotic dynamics, we have seen that the entropy exchange grows at a constant rate $R$ with the number of 
channel uses and we remind the reader that the optimal input state $\hat\rho_{un}$ is left unchanged by the channel so that the coherent information per channel use $I_c/N_q$ can be trivially computed from $R$  by the relation $I_c=N_q-S_e$. Thus in the chaotic case, from (\ref{capacity}) and assuming the channel is degradable, we obtain the capacity $Q=1-R$ which depends exclusively on the coupling strength $\eta$. We can see from fig.(\ref{fig:acop}) that the maximum quantum capacity $Q\to 1$ is reached when there is no coupling $\eta\to 0$ and the capacity tends to its minimum value $Q\to 0$ in the limit of very strong coupling $\eta\gg 1$. On the other hand, when the 
environment's underlying classical dynamics is regular ($-4\leq K\leq 0$)  the entropy exchange only grows logarithmically with the number of channel uses $S_e\propto\log_2 N_q$. This can be seen in Fig.\ref{kpic} where again we have used the 
unpolarized input state $\hat\rho_{un}$ and we plot $S_e$ as a function of $\log_2N_q$ for different values of $K$, all corresponding to regular dynamics. This sub-linear growth of the entropy exchange implies that the channel
is asymptotically noiseless in the case of regular dynamics with resets, $Q=\lim_{N_q\to\infty} I_c/N_q=\lim_{N_q\to\infty} (N_q-S_e)/N_q =1$. 
We can therefore speak of a transition from a noiseless $Q=1$ to a noisy $Q<1$ quantum
channel while varying the classical parameter value $K$ through the transition to  chaos (Fig.\ref{fig:trans}). 
In fact, when $T$ has a finite value, then there will be a crossover of finite width $\eta T$, but in the limit $T\to0$, this crossover becomes a sharp transition (\ref{eq:interaction}). 

In section (\ref{III}) we showed that when the environment's dynamics is chaotic, the coupling parameter also controls the entropy exchange rate (see figure (\ref{fig:acop})), which can vary from $0$ to very close to $1$. Based on the forgetfulness conjecture, we use relations (\ref{eq:info}) and (\ref{capacity}) to suggest that the variation of the $\eta$ can thus be associated with the variation of the channel capacity in the chaotic regime (which is independent of $K$) from noiseless $Q=1$ for zero coupling and 
converging asymptotically to the minimum possible capacity
$Q=0$ for strong coupling. 

\section{Conclusions}\label{VI}
In this article we have investigated a conservative dynamical map as a model of a dephasing quantum channel with few degrees of freedom. We have presented numerical results for the coherent information and entropy exchange in regular and chaotic regime. In the case that the environment's dynamics is chaotic, we have shown the dependence of the entropy exchange rate on a single parameter, namely the interaction strength $\eta$, which plays the role of the single dephasing parameter in stochastic models of dephasing channels. We also considered memory effects in the channel and presented strong physical arguments to support that the present channel model is forgetful in the chaotic regime but not in the regular regime, when memory effects can in general be significant. In order to apply the double-blocking strategy in the case of regular channel dynamics, one must reset the non-chaotic channel and this reset can be efficiently modeled by application of a chaotic map. The model in regular dynamical regime can then also be considered forgetful and encoding theorems can be applied. Based on this assumption, we conjectured a transition from noiseless to noisy channel associated with the transition from regularity to chaos.

We would like to point out that although the results presented here refer to a specific time-dependent interaction Hamiltonian (\ref{eq:interaction}), other time-dependent and time-independent environmental coupling operators $\hat X_E(\hat\theta,\hat{p})$ in eq.(\ref{eq:interaction}) were analyzed and qualitatively equivalent results were obtained in each case, indicating that the specific form of $\hat X_E$ and the fact that it is or not continuous in time does not alter the main features of this quantum dephasing channel model. It is also worth mentioning that we investigated the quantum kicked rotor~\cite{felix} as a model for a dephasing channel. Equivalent results were obtained but as the kicked rotor presents a crossover rather than a transition to chaos, the separate analysis of regular and chaotic regimes presented in this article is more complicated and requires further investigation. 

\begin{acknowledgments}
G.B.L thanks CNPq-Brasil for financial support. We would like to thank very much Antonio D'Arrigo, Filippo Caruso, Cosmo Lupo, Maria Carolina Nemes, Davide Rossini, Fabricio Toscano, and Shashank Virmani for interesting comments on our work and valuable discussions.
\end{acknowledgments}

\end{document}